\documentclass[12pt,preprint]{aastex}

\newcommand\OII{\hbox{[O~{\sc{ii}}}]$\lambda$3727}
\newcommand\OxIII{\hbox{[O~{\sc{iii}}}]$\lambda$5007}
\newcommand\Hbeta{H$\beta$}

\newcommand{\mdot}{M$_{\odot}$~}

\newcommand{\kms}{km~s$^{-1}$}

\shorttitle{Star formation and AGN feedback in HE0450$-$2958}
\shortauthors{Klamer et al.}

\begin{document}

\title{Dressing a {\it{naked}} quasar: star formation and AGN feedback in HE0450$-$2958}


\author{I. J. Klamer\altaffilmark{1}}
\email{Ilana.Klamer@csiro.au}

\author{P. P. Papadopoulos\altaffilmark{2}}
\email{papadop@phys.eth.ch}

\author{R. D. Ekers\altaffilmark{1}}
\email{Ron.Ekers@csiro.au}

\author{E. Middelberg\altaffilmark{1}}
\email{Enno.Middelberg@csiro.au}

\altaffiltext{1}{CSIRO Australia Telescope National Facility, PO Box 76 Epping NSW 1710 Australia}
\altaffiltext{2}{Institut fur Astronomie, ETH, 8093, Zurich, Switzerland}


\begin{abstract}
We present Australia Telescope Compact Array radio continuum observations of the quasar/galaxy system HE0450$-$2958. An asymetric triple linear morphology is observed, with the central radio component coincident with the quasar core and a second radio component associated with a companion galaxy at a projected distance of 7\,kpc from the quasar. The system obeys the far-infrared to radio continuum correlation, implying the radio emission is energetically dominated by star formation activity. However, there is undoubtedly some contribution to the overall radio emission from a low-luminosity AGN core and a pair of radio lobes. Long baseline radio interferometric observations of the quasar core place a $3\sigma$ upper limit of 0.6\,mJy at 1400\,MHz on the AGN contribution to the quasar's radio emission; less than 30\% of the total. The remaining 70\% of the radio emission from the quasar is associated with star formation activity and provides the first direct evidence for the quasar's host galaxy. A re-anlaysis of the VLT spectroscopic data shows extended emission line regions aligned with the radio axis and extended on scales of $\sim$20\,kpc. This is interpreted as evidence for jet-cloud interactions, similar to those observed in radio galaxies and Seyferts. The emission lines in the companion galaxy are consistent with radiative shocks and its spatial association with the eastern radio lobe implies large-scale jet-induced star formation has played a role in this galaxy's evolution.

\end{abstract}



%\section{Introduction- the radio-optical alignment effect}

%Powerful jets, propagating out from the central engine of a galaxy, may interact with the environment along their path. Dense gas clouds encountered as the jets plough through the interstellar medium may be shock-heated and ionised leading to an observed alignment between emission line regions and the axis of radio jets in powerful radio galaxies. In some cases, these jets may even trigger the clouds to collapse and forms stars.  Indeed, even the jets of our closest radio galaxy, Centaurus A (547\,\kms), have shocked, ionised and influenced gas clouds along their path, and even triggered star formation in the intergalactic medium, well beyond the boundary of the host galaxy NGC\,5128. 
%\section{Introduction}

\section{The HE0450$-$2958 system}\label{sourcedes}

HE0450$-$2958 is an optically bright quasar ($M_V=-25.8$) at a redshift of $z=0.285$. The quasar is classified as a Seyfert 1 galaxy, based on its far-infrared (FIR) colours \citep{dgr87} and its optical spectral lines \citep{mer06}. HST imaging of the quasar \citep{boy96,mag05} revealed a 1.7$''$ double optical system containing HE0450$-$2958 and a neighbouring galaxy, separated by a projected distance of 7\,kpc\footnote{We assume a flat $\Lambda$-dominated cosmology with $H_0=75$~\kms Mpc$^{-1}$ and $\Omega_{\Lambda}=0.73$.}. The quasar/galaxy system is a bright IRAS source \citep{dgr87} with a total far-infrared (FIR) luminosity of $L_{\rm FIR}=1\times10^{12}$\,L$_\odot$. Due to the relatively poor spatial resolution of IRAS (5$'$) compared with the small quasar/galaxy separation, it has never been definitively established whether the FIR emission arises from HE0450$-$2958, the neighbouring galaxy, or some combination of both. The total host galaxy light across the system, excluding AGN emission, has been estimated to be about 6L$^*$ \citep{boy96}.

HE0450$-$2958 shot to the forefront of controversy about a year ago when \citet[hereafter M05]{mag05} found evidence for a quasar whose host galaxy was undetected down to a sufficiently interesting limit. M05 estimated a black hole mass of $\approx8\times10^8$~\mdot\ for HE0450$-$2958, by assuming sub-Eddington accretion and 10\% radiative efficiency. Then, using the Magorrian relation between black hole mass and host galaxy bulge luminosity \citep{mag98,mcl02}, they predicted an absolute magnitude of $-23.5\leq M_V\leq-23.0$ for the host galaxy. However, careful deconvolution of the HST images failed to detect a host galaxy associated with HE0450$-$2958 to a limit of $-21.2\leq M_V\leq-20.5$; five to sixteen times fainter than predicted. This intriguing result quickly revived the decade-old debate over the reality of {\it{naked}} quasars \citep{bah94}, leading to a flurry of theoretical papers describing various scenarios for ejecting black holes from their host galaxies during violent mergers \citep{hof06}, and the implications of such violent events for future gravity wave detectors  \citep{hae06}. More recently \citet{mer06} argued that HE0450$-$2958's black hole mass could be an order of magnitude smaller, leading to a host galaxy luminosity of $M_V\approx-21$, and therefore quite consistent with the upper limit derived from the HST images. 

M05 attributed all of the FIR emission to thermal radiating dust grains associated with the companion galaxy 7\,kpc from HE0450$-$2958. However, there remains the possibility of a dusty (optically obscured) host galaxy associated with HE0450$-$2958\footnote{A nearby analog of an optically obscured host galaxy is Markarian 231, a Seyfert 1 quasar with a massive, compact (radius $\rm \sim  450\,pc$) molecular disk fueling a spectacular, but heavily obscured, starburst \citep{dow98}.}. Radio continuum emission from galaxies is a good tracer of star formation, and is not obscured by dust. To this end, we observed HE0450$-$2958 with the Australia Telescope Compact Array (ATCA). In this paper, the results of the ATCA imaging campaign are presented, along with a re-analysis of archival VLA and VLT data. We find that whilst the radio emission from the HE0450$-$2958 system is energetically dominated by star formation activity, a fraction of the radio emission must be driven by a pair of low-luminosity radio lobes and a radio core. Throughout this paper we define the HE0450$-$2958 system to mean the quasar/galaxy pair, and unless otherwise stated we refer to HE0450$-$2958 itself as simply {\it {the quasar}}.

\section{Observations \& Results}\label{results}
\subsection{ATCA}\label{ATCA}
We observed a field centred on HE0450$-$2958 with the ATCA during 2006 April. The standard continuum correlator configuration was used which consists of two frequency windows, each with 128\,MHz of bandwidth separated into 32 spectral channels, allowing dual frequency observations. We chose RFI-free parts of the 6 and 3\,cm bands and centred our frequency windows at 6208 and 8640\,MHz accordingly. The most extended array configuration was used which has a maximum baseline of 6\,km and translates into a spatial resolution of order $4''\times2''$ at $-29^{\circ}$ declination.  

HE0450$-$2958 was observed for 12 hours on two separate occasions. Flux density calibration was achieved to within 5\% uncertainty, based on a short observation of the ATCA primary flux calibrator PKS\,B1934$-$638. We corrected our observations for atmospheric phase fluctuations by intermittedly observing the strong, unresolved quasars PKS\,B0451$-$282 and PKS\,B0454$-$234 each for 2 minutes every other 30 minutes throughout the 12 hour observations. Data analysis was performed using the standard \textsc{miriad} procedures for centimetre observations as outlined in the users' guide\footnote{http://www.atnf.csiro.au/computing/miriad}. Due to non-negligible phase instability during at the beginning and end of each observing run, we flagged the earliest and latest hour angles on each day, resulting in a fairly elongated synthesised beam. The observing parameters are given in Table~\ref{journal}. The signal-to-noise of the data are not sufficient to reliably explore polarisation structure. 

\clearpage
\begin{deluxetable}{ccccc}
\centering\tabletypesize{\scriptsize}
\tablecaption{Observational parameters\label{journal}}
\tablewidth{0pt}
\tablehead{
\colhead{Frequency} & \colhead{$\sigma_{rms}$$^a$} & \colhead{$\theta \times \theta$$^b$} &  \colhead{PA$^b$} & \colhead{FOV$^c$} \\
\colhead{MHz} & \colhead{mJy~beam$^{-1}$} & \colhead{$''\times''$}  & \colhead{$^{\circ}$} & \colhead{$'$}
}
\startdata

6208 & 0.04 & 5.68$\times$1.95 & 16.1 & 7.7\\
8640 & 0.05 & 4.07$\times$1.25 & 14.9 & 5.5\\

\enddata
\tablecomments{$^a$rms noise level around the source in the deconvolved image. $^b$Size and position angle of the restoring beam. $^c$Full width at half maximum (FWHM) of the primary beam.}
\end{deluxetable}

\begin{deluxetable}{ccccccccc}
\tabletypesize{\scriptsize}
\tablecaption{Radio flux densities\label{fluxes}}

\tablewidth{0pt}
\tablehead{
\colhead{Component} & \colhead{RA} &\colhead{DEC} &\colhead{S(843)$^a$} & \colhead{S(1400)$^b$} &\colhead{S(6208)}&\colhead{S(8640)}&\colhead{$\alpha^{8640}_{6208}$}&\colhead{$\alpha^{1400}_{843}$}\\
&$^{h\,\,m\,\,s}$&$^{h\,\,m\,\,s}$&\colhead{mJy}&\colhead{mJy}&\colhead{mJy}&\colhead{mJy}\\
}
%\colhead{Component} & \colhead{S(843)} &\colhead{S(1400)} &\colhead{S(6208)} & \colhead{S(8640)}\\
%\colhead{} & \colhead{mJy} &\colhead{mJy} &\colhead{mJy} & \colhead{mJy}  \\
%}
\startdata
C1 (quasar) & 04 52 30.1 & -29 53 35.34 & \nodata & \nodata & $1.8\pm0.05$ & $1.2\pm0.05$ & $-1.2\pm0.2$ & \nodata \\
C2 (companion) & 04 52 30.2 & -29 53 36.27 & \nodata & \nodata & $1.0\pm0.05$ & $0.8\pm0.05$& $-0.8\pm0.4$ & \nodata\\
C3 & 04 52 30.0 & -29 53 35.26 & \nodata & \nodata & $0.4\pm0.05$ & $0.3\pm0.05$ & $-0.9\pm0.8$ & \nodata\\
\\
integrated & \nodata & \nodata & $17.0\pm4.3$ & $9.5\pm 0.5$ & $3.4\pm 0.05$ & $2.3\pm 0.04$ & $-1.3\pm0.2$ & $-1.1\pm0.7$
\enddata
\tablecomments{$^a$The 843\,MHz SUMSS \citep{mau03} flux density is measured with a spatial resolution of 84$''\times$45$''$. $^b$The 1400\,MHz \citep{con98} NVSS flux density is measured with a spatial resolution of 45$''\times$45$''$.}
\end{deluxetable}
\clearpage

\begin{figure}\centering
\includegraphics[width=11cm]{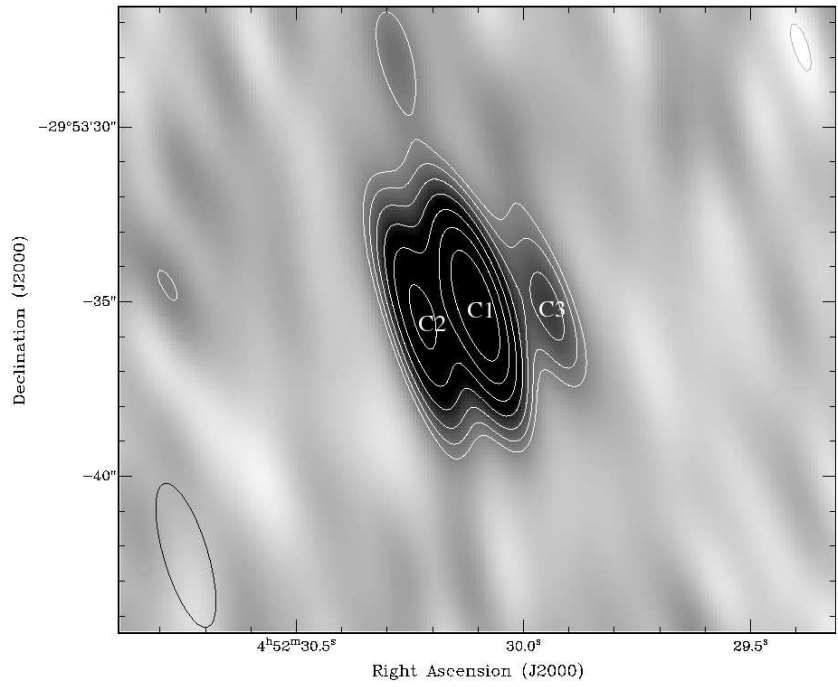}
\caption{ATCA 6208\,MHz image (robust=0 tapering) of the HE0450$-$2958 system with contours of the same image overlaid. The three resolved components C1, C2, C3 are labelled. The size of the restoring beam is given in the bottom left corner. The contour levels are $\pm(150, 212, 300, 424, 600, 849, 1200)\,\mu$Jy\,beam$^{-1}$. }\label{ATCAplots}
\end{figure}

Figure~\ref{ATCAplots} shows the 6208\,MHz radio continuum greyscale image produced with a robust=0 weighting applied to the visibilities, overlaid with flux-density contours of the same data. A triple linear structure is resolved: the brightest component (labelled C1) in the centre, a second component to the east (C2) and a third component to the west (C3).

\begin{figure}
\centering
\includegraphics[width=11cm]{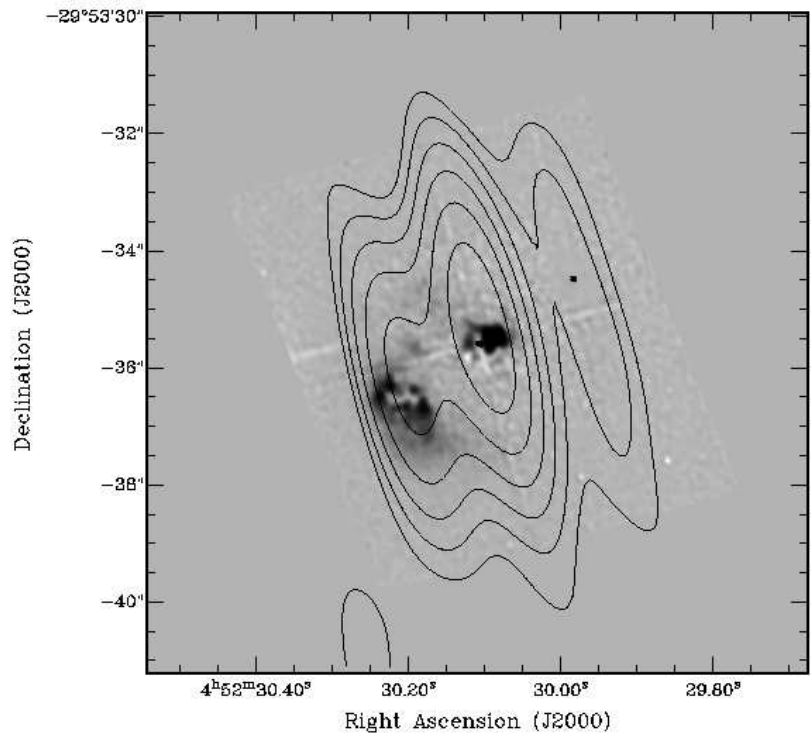}
\caption{PSF-subtracted HST image from M05 overlaid with flux density contours of the naturally weighted 8640\,MHz image. The contour levels are the same as those given in Figure~\ref{ATCAplots}. The quasar and star, shown as point sources in the image, are saturated.}
\label{HSTimage}
\end{figure}

Figure~\ref{HSTimage} shows the astrometrically corrected and Point Spread Function (PSF) subtracted HST image published by M05. The astrometric accuracy of the image is in the range 0.5$''$-1.0$''$ (P. Magain, private commun.). The quasar is the central point source and it is surrounded by diffuse ionised emission with no strong continuum detected. The high surface brightness region about 0.5$''$ west of the quasar --- aptly named the {\it{blob}} --- is also dominated by ionised emission lines and has no strong continuum. There is a foreground G-type star located north-west of the quasar, and a complicated looking companion object with the spectral characteristics of a starforming galaxy, to the south-east. Overlaid on the HST image are the flux-density contours of the 8640\,MHz radio continuum image produced with natural weighting of the visibilities. Clearly, C1 is coincident with the quasar and C2 is associated with the companion galaxy, although it it is appears offset (in projection) by about 0.5$''$. Although C3 lies remarkably close in projection to the foreground star, it is almost certainly unrelated to it. Stellar radio emission is very rare and typically only detected from hot OB associations \citep{bie89} and symbiotic stars \citep{sea84}. 

\clearpage

\clearpage

\subsection{VLA}\label{vla}
HE0450$-$2958 was observed with the Very Large Array (VLA) for 9 minutes during 1990 July 6 in the AB configuration at 1500, 4800 and 8400\,MHz (AH0406). We retrieved these data from the VLA archive and analysed them using the \textsc{runvla} task which is now implemented in the \textsc{aips} data reduction software. The task is a pipeline routine written by Lorant Sjouerman (NRAO, Socorro) for the purpose of reducing VLA archive data.

The triple structure observed in the ATCA images is unresolved in the VLA 1500\,MHz image (angular resolution of $6''\times4''$) but the integrated flux density is consistent with the flux density in the NVSS catalogue \citep{con98} and therefore constrains variability on 3--6 year timescales. At 8400\,MHz, only the quasar is confidently detected, with a flux density of ($0.6\pm0.2$)\,mJy, compared to the ATCA flux density of ($1.2\pm0.05$)\,mJy for the quasar (C1). The discrepancy between the VLA and ATCA flux densities can be explained by the factor of seven difference in the spatial resolution between the two datasets, and immediately implies that 50\% of the radio continuum emission at the position of the quasar is extended on spatial scales of greater than about 3\,kpc. Such extended emission is unlikely to be due to AGN emission, and is most likely from star formation activity associated with the host galaxy; we return to this in \S\ref{starformationargument}.

\begin{figure}\centering
\rotatebox{270}{\includegraphics[width=11cm]{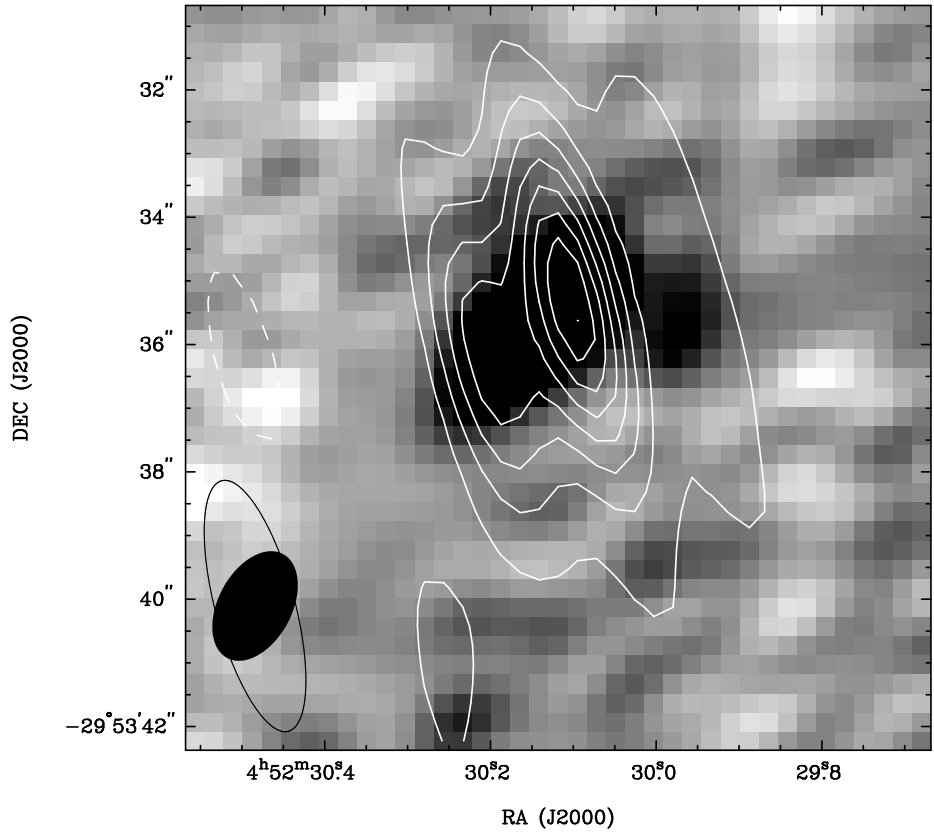}}
 \caption{VLA 4.8\,GHz greyscale image overlaid with ATCA 8640\,MHz contours. The contour levels are the same as those given in Figure~\ref{ATCAplots}. The VLA restoring beam size is shown as the filled ellipse inside the ATCA beam in the bottom left corner.}\label{vlaimage}
\end{figure}

The 4800\,MHz VLA data (shown in greyscale in Figure~\ref{vlaimage}) verify the triple radio structure observed in the ATCA images; C1 and C2 are not separately resolved, but C3 is resolved with a flux density of ($230\pm115$)\,$\mu$Jy~beam$^{-1}$. The poor signal-to-noise of the data preclude any further analysis of this dataset. Deeper VLA observations will be crucial for us to further to explore the spatial distribution of the radio emission in detail. 

\clearpage
\clearpage
\subsection{VLT spectroscopy}
The HE0450$-$2958 system was observed with the ESO Very Large Telescope (VLT) in November 2000, using the FORS1 instrument in Multi-Object Spectroscopy (MOS) mode with three medium resolution grisms (G600B, G600R and G600I) and covered the wavelength range from 3500-9000\AA. One slitlet was centred on the quasar itself, and the slit position angle was aligned along the star-quasar-galaxy axis shown in Figure~\ref{HSTimage}. In order to separate the quasar nucleus from the diffuse surrounding emission, a PSF correction was applied to the dataset using a method developed for the spatial deconvolution of spectra \citep{cou00}, based on the MCS deconvolution technique for images \citep{magain98}. Spectroscopic processing details are given in M05.

\section{Star formation in HE0450$-$2958}\label{starformationargument}
What is the nature of the observed radio continuum emission? We have detected a kilo-parsec scale triple linear radio structure centred on a quasar. Is it dominated by AGN-related emission (core, jets, lobes, hotspots) issuing out from the quasar itself? Or, given the large FIR luminosity associated with the HE0450$-$2958 system, could the radio emission be energetically dominated by star formation activity from both the quasar's host galaxy and the nearby companion? And, if the system is dominated by star formation, then what is the origin of the western radio component (C3), not associated with any optical counterpart?  

In \S\ref{FIRRCcor}-\S\ref{vlbi}, we will show that the HE0450$-$2958 system obeys the far-infrared to radio continuum correlation for normal (i.e. non-AGN) galaxies, and that the quasar's radio emission has a steep spectral index and is resolved out in long-baseline interferometry data. Together, these three conditions strongly favour the radio emission from the HE0450$-$2958 system being dominated by star formation activity. 

\subsection{The far-infrared to radio continuum correlation}\label{FIRRCcor}

\begin{figure}
\centering\includegraphics[width=12cm]{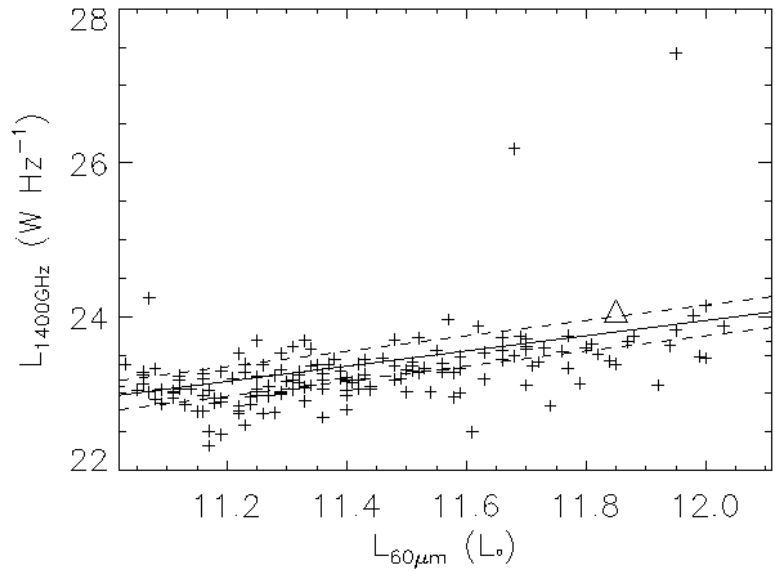}
\caption{The FIR-RC correlation for the subset of galaxies in Y01 with log $L_{60\mu \textrm m}>11\;L_{\odot}$ (crosses) and for HE0450$-$2958 (triangle). The solid and dashed lines correspond to the formal fit to the FIR-RC correlation given in Equation~\ref{FIRRC}. The three sources lying well above the FIR-RC correlation are examples of sources in the IRAS 2\,Jy sample with an excess of radio continuum emission from an AGN. }\label{yun2001figure5}
%\centering\includegraphics[width=8cm]{plotq.eps}
%\caption{The q parameter for the subset of galaxies in Y01 with log $L_{60\mu \textrm m}>11\;L_{\odot}$ (crosses) and for HE0450$-$2958 (triangle). The solid and dashed lines represent the range $\bar{q}\,\pm\,\sigma_q$ from Y01 (see text for details). The three sources with $q\leq1.64$ have an excess of radio continuum emission from an AGN.}
\label{yun2001figure6}
\end{figure}

There is a very tight correlation between the far-infrared (FIR) and radio continuum (RC) emission from normal (ie. non-AGN) galaxies \citep{con92}. While the actual mechanism driving the FIR-RC correlation is still uncertain, this does not diminish the importance of the empirical relationship. It is generally accepted that the fundamental ingredient is star formation: the FIR component is due to thermal radiation from dust grains which absorb the optical/far-UV starlight and re-radiate it in the infrared regime, whereas the radio continuum component arises from magnetic fields with synchrotron radiating relativistic electrons related to the star formation activity (see also Figure~\ref{yun2001figure5}). 

If the HE0450$-$2958 system obeys the FIR-RC correlation, there is a good chance that most of its radio continuum emission is dominated by star formation processes, rather than more powerful AGN emission from the central quasar. As mentioned in \S\ref{sourcedes}, due to the relatively low angular resolution of the existing IR observations (from IRAS), it is currently not possible to separate the FIR component associated with the quasar position from that associated with the neighbouring galaxy. However, we can constrain the nature of the quasar emission by measuring the location of the integrated quasar/galaxy system on the FIR-RC correlation (also see \citealp{cra92,roy98}). 

We use the form of the FIR-RC correlation derived from 1809 infrared selected galaxies from the IRAS 2\,Jy catalogue\footnote{It makes sense to compare HE0450$-$2958 with an IRAS selected sample because HE0450$-$2958 was itself infrared selected.} by \citet[hereafter Y01]{yun01}, 

\begin{equation}
\textrm {log}\;L_{1400}=(0.99\pm0.01)\;\textrm {log}\;L_{60\mu \textrm m}\,+(12.07\pm0.08),
\label{FIRRC}\end{equation}
where 
\begin{equation}
%\text {log}\frac{L_{1400}}{\text W\text m^{-2}\text {Hz}^{-1}}= 20.08 + 2\;\text{log}\;\frac{D}{\text {Mpc}} +\text{log}\;\frac{S_{1.4}}{\text{Jy}},
\textrm {log}\frac{L_{1400}}{\textrm W\textrm m^{-2}\textrm {Hz}^{-1}}= 20.08 + 2\;\textrm{log}\;\frac{D}{\textrm {Mpc}} +\textrm{log}\;\frac{S_{1.4}}{\textrm{Jy}},
\label{nvss}\end{equation}
and
\begin{equation}
\textrm {log}\frac{L_{60\mu \textrm m}}{L_{\odot}} = 6.014 +2\;\textrm {log}\;\frac{D}{\textrm {Mpc}} + \textrm {log}\;\frac{S_{60\mu \textrm m}}{\textrm {Jy}},
\end{equation}

 where D is the source distance. Substituting $S_{60\mu \textrm m}=(650\pm52)$\,mJy (De Grijp et al. 1987) into Equation~\ref{FIRRC} yields $\textrm {log} \; L_{1400}=(23.80\pm0.23)$. This is consistent --- but at the lower limit --- with the observed value of $24.08\pm0.02$ from Equation~\ref{nvss} with $S_{1400}=(9.5\pm0.5$)\,mJy from the NVSS catalogue \citep{con98}. The position of the integrated HE0450$-$2958 quasar/galaxy system on the FIR-RC correlation for a subset of the most FIR luminous galaxies from Y01 is shown in Figure~\ref{yun2001figure5}. 

%A second, equivalent measure of the compatibility of the quasar/galaxy system with the FIR-RC correlation is the $q$ parameter, shown in Figure~\ref{yun2001figure6}. The $q$ parameter is a logarithmic ratio of the FIR to RC flux-density, as first defined by \citet{hel85}, 
%\begin{equation}
%q\equiv \rm log\frac{FIR}{3.75\times10^{12}}-log\frac{S_{1.4}}{10^{26}}.
%\label{qequation}\end{equation}
%Y01 derive a mean $\bar{q}= 2.34$ with a dispersion $\sigma_q=0.33$ for far-infrared sources with $L_{60\mu \text m}>10^{11}\;L_{\odot}$. For the HE0450$-$2958 system, we measure $q=1.94^{+0.06}_{-0.07}$, consistent with having its radio continuum emission dominated by star formation processes. Radio-excess objects, {\it {i.e.}} sources with additional radio emission from an AGN, are defined to have $q\leq1.64$ (Y01). The HE0450$-$2958 system clearly does not fit the definition of radio-excess.
\clearpage
\clearpage

Since  53\% of the radio continuum emission from the HE0450$-$2958 system originates at the position of the quasar, and since the quasar/galaxy system obeys the FIR-RC correlation, this suggests that approximately 53\% of the FIR emission arises at the quasar position, and is powered by star-forming activity from its surrounding host galaxy. This could explain why HST failed to detect the host galaxy: it is simply heavily obscured by the dust from a luminous ($L_{\rm FIR}=5\times10^{11}$\,M$_\odot$) infrared galaxy, with an implied star formation rate of order 50\,\mdot\,year$^{-1}$ \citep{con92}. This seemingly large star formation rate is typical for locally observed luminous infrared galaxies \citep{dun00} and infrared bright quasars \citep{cra92}. 

Due to the small, but finite, dispersion in the FIR-RC correlation, a scenario in which some fraction of the radio continuum emission from the HE0450$-$2958 system is due to AGN activity is certainly possible; it is not uncommon to find AGN and star formation activity in the same system. However, AGN emission cannot be the dominant process contributing to the radio continuum emission of the HE0450$-$2958. 
\subsection{Spectral Index considerations}\label{alpha}

Table~\ref{fluxes} lists the values of $\alpha^{8640}_{6208}$ for C1 (quasar), C2 (companion) and C3. A combination of low signal-to-noise and proximity of the two frequencies means that our estimates of $\alpha^{8640}_{6208}$ for both C2 and C3 have large associated uncertainties. However, we have measured $\alpha=-1.2\pm0.2$ for C1 (the quasar), which does rule out a flat-spectrum AGN component (typically the radio continuum emission from a compact AGN core is flatter than $-0.5$). The observed spectral index is consistent with optically thin synchrotron emission either from star formation or extended low-luminosity AGN structures like compact steep-spectrum jets and/or lobes. Steep radio spectral indices are typical of the cores of Seyfert galaxies (for example \citealp{deB78,ulv84,ulv89,mor99,rus96}).
\subsection{Long baseline interferometry}\label{vlbi}

HE0450$-$2958 was included in a sample of Seyfert galaxies observed with the now defunct 275\,km single baseline Parkes-Tidbinbilla real-time interferometer (PTI; Norris et al. 1988, Roy et al. 1994). At 2300\,MHz, the PTI was only sensitive to structures smaller than 0.1$''$, and therefore perfectly suited to probe compact radio continuum emission from the nucleus of Seyferts and quasars. HE0450$-$2958 was observed with the PTI (two snapshots of the {\textit{uv}}-plane) in 1987, 1990 and 1991 \citep{roy94,roy98}. The quasar was {\it not} detected down to $3\sigma=2$\,mJy at 2300\,MHz. Using the observed radio spectral index of $\alpha^{6208}_{8640}=-1.2\pm0.2$ (see \S\ref{results}), this means that $<$0.6\,mJy of the radio continuum emission at 6208\,MHz is contaminated by an AGN core; this upper limit is a factor of three smaller than the total radio continuum emission observed from C1. The lack of compact radio continuum emission down to these limits implies that C1 is resolved on scales of $\gtrsim400$\,pc. Coupled to the good fit to the FIR-RC correlation (\S\ref{FIRRC}) these long baseline constraints also support a star formation origin for the bulk of the quasar's radio continuum emission, arising from its underlying host galaxy.

\section{AGN core and jet emission}\label{agn}
The bulk of the radio continuum emission from the HE0450$-$2958 system arises from star formation activity at the position of the quasar (C1) and the companion galaxy (C2). This in itself is an interesting result: {\it {it is direct evidence for a host galaxy associated with the quasar}}. However, the ATCA radio images (confirmed by the VLA archive images) show an asymetric, almost-linear, triple radio structure with a projected size of $\sim$20\,kpc. Although C1 and C2 can both be identified with optical counterparts, and therefore understood in terms of star formation,  C2 is not perfectly coincident with the companion galaxy, and C3 has no associated optical counterpart. A full description of the HE0450$-$2958 system must include C3 and an explanation for the slight offset between C2 and the companion, so there is something plainly missing from the pure star formation interpretation. The linearity of the three radio components suggests the presence of low-luminosity jets similar to those observed in Seyfert galaxies, or radio-quiet quasars. In this case, the AGN core is centred on C1 (the quasar), and C2 and C3 would be low-luminosity radio lobes, or poorly resolved jets.

In \S\ref{FIRRCcor}, we mentioned that whilst the total radio continuum emission is dominated by star formation, the system would still obey the FIR-RC correlation if a fraction was due to AGN processes. The integrated radio continuum emission of the system is 3.4\,mJy at 6208\,MHz. At this frequency, C3 is associated with 0.4\,mJy of flux density which we now attribute to the western radio lobe. We associate an equal contribution of 0.4\,mJy to lobe emission at C2. In addition, a maximum 0.6\,mJy can be associated with AGN core emission from C1 (see \S\ref{vlbi}). In total, up to 41\% of the total radio continuum emission from the system can be attributed to an AGN contribution. This maximum estimate translates into a 1400\,MHz radio luminosity of order $1\times10^{24}$\,W\,Hz, typical of low-luminosity Seyfert or FR\,I radio sources. %We note that the flux density of C2's lobe emission could be higher than C1's, since the conversion from quasar jet power to radio luminosity is more efficient in higher density media (e.g. \citealp{bart96}) and C2 is located in the vicinity of a galaxy, whereas C3 is in the intergalactic medium. 
%\section{AGN feedback from the quasar}
\section{Extended emission line regions}

\begin{figure}
\centering\includegraphics[width=12cm]{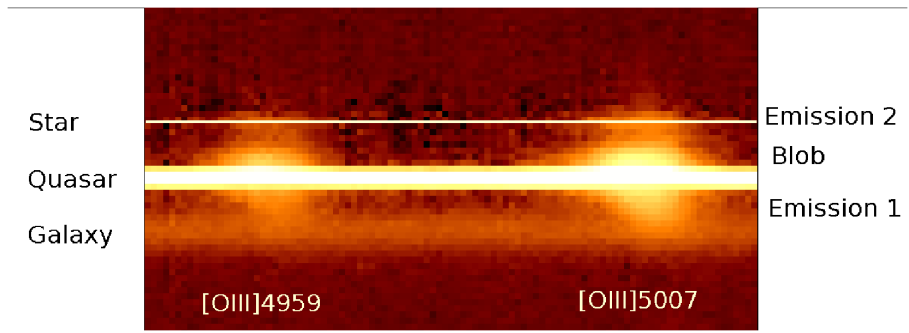}
\caption{The 2D VLT spectra from M05, around the \OxIII\ doublet emission lines. See the text for details.}\label{2Dspec}
\end{figure}

%\subsection{Extended emission line regions}
We have reanalysed the VLT spectroscopic data presented in several previously published papers (M05, \citealp{mer06}). Figure~\ref{2Dspec} shows the 2D spectra around the \OxIII\ lines, where the point sources (star and quasar) have been subtracted from the original spectra and then artificially added with an increased resolution at the appropriate position; this was done to disentangle the various diffuse components. As expected, the star, galaxy and quasar all show continuum emission. However, there are also at least three separate emission line regions with no detectable continuum: one between the quasar and galaxy (labelled {\it{Emission 1}} in Figure~\ref{2Dspec}), one between the star and the quasar ({\it{Emission 2}}), and a high surface brightness region (the blob) about 0.5$''$ west of the quasar. {\it{Emission 1}} and the blob are both detected in the HST optical image shown in Figure~\ref{HSTimage}, but {\it{Emission 2}} is not detected presumably due to the ACS sensitivity (Letawe, private communication). Velocity structure is quite evident in the spectral profile of {\it{Emission 1}}.
 %However, the foreground star was assumed to be a point source and used to constrain the PSF of the quasar. This means that some of the low surface brightness extended emission could have been included in the PSF and therefore removed in the deconvolution process. The latter implies the diffuse emission seen in the HST image could be missing some low-surface brightness component.

Extended emission line regions are typically found aligned along the radio axes of powerful radio galaxies \citep{mcc87,cha87,mcc93,wvb98}. In many cases there is very good evidence that the alignment is caused by the interaction of the radio jets with gas clouds in the interstellar medium; the jets shock-ionise the clouds along their path (e.g. \citealp{dic95,fos98,tad00}) and in some cases induce gravitational cloud collapse, which can trigger star formation \citep{dey97,wvb93,cro06}. In other cases, the emission line regions are dominated by photoionisation by the AGN (illumination). No doubt it is often a combination of all of these that produce the alignment effects we observe.  

\begin{figure}
\centering\includegraphics[width=12cm]{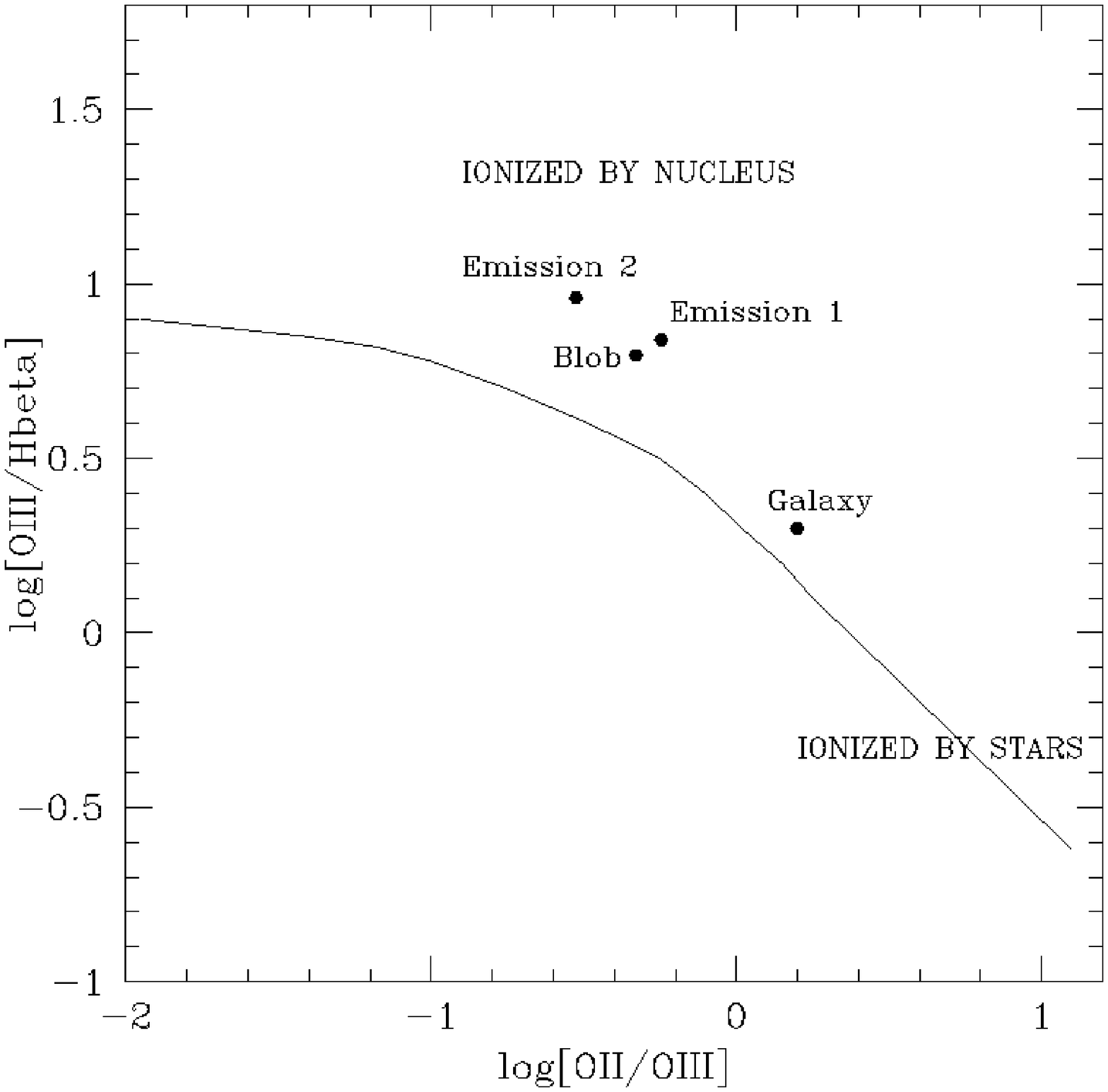}
\caption{The \OxIII/\Hbeta\ versus \OII/\OxIII\ line ratios for the various sources in the HE0450$-$2958 system. The solid line is the \citet{bal81} fit for photoionisation by HII regions. This figure indicates that the blob, quasar and emission line regions 1 and 2 are photoionised by the AGN, whereas the companion galaxy lies within the parameter space of pure shock-heated galaxies, albeit at the lower limit and close to the HII region line.}
\label{diagnostic}
\end{figure}

In Figure~\ref{diagnostic}, a BPT \citep{bal81} diagnostic diagram has been constructed showing the \OxIII/\Hbeta\ versus \OII/\OxIII\ line ratios for the various emission line regions shown in Figure~\ref{2Dspec}. Such a diagram is a useful way to separate photoionisation by AGNs from photoionisation by HII regions; AGNs have systematically larger \OxIII\ line intensities for a given \OII/\OxIII\ ratio. Clearly the blob, and regions 1 and 2 from Figure~\ref{2Dspec} are being ionised by AGN activity and not star formation. However, it is notoriously difficult to use diagnostic diagrams to actually separate AGN illumination from ionisation by radiative shocks from AGN jets \citep{dop95}, because they occupy a similar parameter space in the diagrams (see \citealp{ins06,tad00} for recent some diagnostic diagrams of the extended emission line regions around radio galaxies). We are, therefore, unable to confidently distinguish between the AGN illumination and shock ionisation based on the diagnostic plot alone. However, given the good spatial coincidence between the radio structure and the emission line clouds, we favour the {\it {shock+precursor}} ionisation scenario triggered by jet-cloud interactions \citep{dop95}. Higher resolution radio images, along with optical integral field spectroscopy would be an excellent way to to explore the spatial distribution of the jets/lobes with the emission line gas in more detail in order to confidently determine the origin of the radio-optical alignment effect in the HE0450$-$2958 system.

\clearpage
\clearpage 

What about the companion galaxy, which appears to be (in projection) physically associated with the eastern radio lobe? In Figure~\ref{diagnostic} the companion galaxy is located within the parameter space of pure shock-heated galaxies \citep{bal81,dop95}, albeit at the lower limit and close to the HII region line. It is likely that shocks created by the passage of the radio jet/lobe can explain the location on the diagnostic diagram. The physical coincidence between the radio lobe and the companion galaxy is strong evidence for large scale jet-induced star formation. This would be analogous to the star formation recently detected in the lobes of Seyfert galaxies \citep{xan06} and the discovery of a SCUBA source coincident with the lobe of a powerful radio galaxy \citep{ivi02}. The higher resolution observations we proposed above, are of course essential to verify a jet-induced star formation scenario for the companion galaxy.

%\section{Interpreting the HE0450$-$2958 system}
%The HE0450$-$2958 quasar is absolutely typical of Seyfert galaxies without radio cores \citep{roy98}. In these sources, the radio emission is usually extended on kpc scales, the radio spectral indices are steep, the systems obey the FIR-RC correlation, and the radio emission is assumed to be due to some combination of star formation and AGN activity.

\section{Summary}
The main conclusions of this paper are:

\begin{enumerate}
\item{Using the Australia Telescope Compact Array, we have resolved a triple linear radio continuum structure associated with the HE0450$-$2958 quasar/companion galaxy system at $z=0.285$.}
\item{The radio emission is dominated by star formation activity from the quasar's host galaxy. This is the first evidence for a host galaxy in the HE0450$-$2958 quasar, and contradicts any suggestion that it is a `naked' quasar.}
\item{We also find clear evidence for low-luminosity radio lobes, analogous to those of Seyferts or radio-quiet quasars.}
\item{The brightest radio component is coincident with the quasar, and is energetically dominated by star formation activity from its associated host galaxy. Some fraction of the radio continuum emission may be associated with an AGN core, but this is less than 30\% of the total radio emission detected at the position of the quasar. }
\item{The second brightest radio component is physically associated with the companion galaxy located 7\,kpc from the quasar. The emission is a combination of a low-luminosity (eastern) radio lobe, typical of Seyfert or radio-quiet quasars, and star formation activity from the companion galaxy itself.}
\item{The third, and faintest, radio component has no associated optical counterpart, and is interpreted as the counter (western) radio lobe. }
\item{Highly ionised optical emission line regions are resolved along the radio axis, extending across the entire radio structure. The ionisation mechanism is most likely due to radiative shocks by the passage of the propagating jets that interact with existing gas clouds in the interstellar/intergalactic medium. However, a part of the extended emission line region could be AGN illumination from the quasar core.}
\item{The companion galaxy is physically associated with the eastern radio lobe, shows clear signs of star formation activity, and has ionised emission lines consistent with pure shock models. Taken together, we favour a scenario in which the companion is the result of large-scale jet induced star formation.}\\
\end{enumerate}

High resolution, deep VLA and VLBI radio observations are needed to explore the nature of the radio emission in more detail. Spectropolarimetry and optical/near-IR integral field spectroscopy on the extended emission line regions will constrain the ionisation mechanism (be it AGN ilumination or shocks from jet-cloud interactions). Planned high resolution observations of the CO J=1--0 emission line in the system are expected to shed more light on the distribution of the large expected molecular gas reservoir fuelling the star formation activity in the quasar host and the companion galaxy.

\acknowledgments
The Australia Telescope Compact Array is part of the Australia Telescope which is funded by the Commonwealth of Australia for operation as a National Facility managed by CSIRO. The authors are extremely grateful to Pierre Magain and Geraldine Letawe for providing us with their optical HST image and the 2D VLT spectra, for creating Figures~\ref{2Dspec} and \ref{diagnostic}, and for extremely useful discussion regarding this manuscript. A big thank-you also goes to Lorant Sjouerman for his help reducing the VLA archive data. PPP thanks Marcela Carollo for bringing the HE0450$-$2958 system to his attention; this was instrumental in initiating this project.

%\bibliography{mnemonic,mnemonic-simple,biblio}
\end{document}